\documentclass[conference]{IEEEtran}
\IEEEoverridecommandlockouts
\usepackage{cite}
\usepackage{amsmath,amssymb,amsfonts}
\usepackage{graphicx}
\usepackage{float}
\usepackage{subfig}
\usepackage{textcomp}
\usepackage{xcolor}
\usepackage{csquotes}
\usepackage{multirow}
\usepackage{tikz}
\usepackage{siunitx}
\usetikzlibrary{shapes.geometric, arrows}
\usepackage{makecell}
\usepackage{float} 
\usepackage{flushend}
\usepackage{enumitem}
\usepackage{booktabs}
\usepackage{makecell}
\usepackage{multirow}
\usepackage{titlesec}
\usepackage{xspace}
\xspaceaddexceptions{)}
\usepackage{glossaries}
\usepackage{array}
\usepackage{todonotes}
\usepackage{booktabs}

\newglossaryentry{vdd}{
	name={\ensuremath{V_{\text{DD}}}},
	description={supply voltage},
	symbol=Vdd,
	sort=vdd}
\newglossaryentry{vth}{
	name={\ensuremath{V_{\text{TH}}}},
	description={threshold voltage},
	symbol=Vth,
	sort=vth}
\newglossaryentry{gnd}{
	name={GND},
	description={El. ground},
	symbol=GND}

\newacronym{tcad}{TCAD}{technology CAD}

\newacronym{fefet}{FeFET}{ferroelectric FET}
\newacronym{fe}{FE}{ferroelectric}
\newacronym{ncfet}{NCFET}{negative capacitance FET}
\newacronym{feram}{FeRAM}{ferroelectric RAM}

\newacronym{fdsoi}{FD-SOI}{fully depleted silicon on insulator}
\newacronym{tcam}{TCAM}{Ternary Content Addressable Memory}

\newacronym{sap}{SAP}{stuck-at-polarization}

\newacronym{beol}{BEOL}{back-end-of-line}
\newacronym{feol}{FEOL}{front-end-of-line}
\newacronym{mfm}{MFM}{metal-ferroelectric-metal}

\newacronym{po}{PO}{primary output}
\newacronym{pi}{PI}{primary input}
\newacronym{cf}{CF}{criticality-factor}
\newacronym{mlp}{MLP}{multi-layer-perceptron}
\newacronym{atpg}{ATPG}{Automatic Test Pattern Generation}
\newacronym{ic}{IC}{integrated circuit}
\newacronym{htd}{HTD}{hard-to-detect}
\newacronym{wfp}{WFP}{weighted fault-propagation}
\usepackage{cleveref}
\crefname{enumi}{Step}{Steps}
\crefname{section}{Section}{Sections}
\crefname{subsection}{Section}{Sections}
\crefname{figure}{Figure}{Figures}
\crefname{table}{Table}{Tables}
\crefname{equation}{Equation}{Equations}
\crefname{eqnarray}{Eq.}{Eq.}
\crefname{appendix}{Section}{Sections}
\usepackage{algorithm}
\usepackage{algpseudocode}
\usepackage{amsthm}
\usepackage{array}
\usepackage{tabularx}
\usepackage{etoolbox}
\usepackage{xcolor}
\usepackage{soul}
\sethlcolor{yellow}
\usepackage[a4paper, total={184mm,239mm}]{geometry}
\makeatletter
\def\subsubsection{\@startsection{subsubsection}{3}{\z@}%
  {0ex plus 0.1ex minus 0.1ex}{0ex}%
  {\normalfont\normalsize\bfseries}}

\def\@IEEEsectpunct{:\ }
\makeatother

\newcolumntype{Y}{>{\centering\arraybackslash}X}

\newtheorem{lemma}{Lemma}
\algrenewcommand{\algorithmiccomment}[1]{\hfill// \textit{#1}}
\renewcommand{\footnoterule}{\kern-3pt\hrule width 0.4\columnwidth\kern 2.6pt}
\tikzstyle{startstop} = [rectangle, rounded corners, 
minimum width=3cm, 
minimum height=1cm,
text centered, 
text width=3.5cm, 
draw=black, 
fill=gray!30]
\tikzstyle{process} = [rectangle, 
minimum width=3cm, 
minimum height=1cm, 
text centered, 
text width=3cm, 
draw=black, 
fill=gray!30]
\tikzstyle{ml_model} = [rectangle, 
minimum width=3cm, 
minimum height=2cm, 
text centered, 
text width=2.5cm, 
draw=black, 
fill=white!100]
\tikzstyle{io} = [trapezium, 
trapezium stretches=true, % A later addition
trapezium left angle=70, 
trapezium right angle=110, 
minimum width=3cm, 
minimum height=1cm, 
text centered, 
text width=3cm, 
draw=black, 
fill=gray!30]
\tikzstyle{arrow} = [thick,->,>=stealth]

\newcolumntype{P}[1]{>{\centering\arraybackslash}p{#1}}

\def\BibTeX{{\rm B\kern-.05em{\sc i\kern-.025em b}\kern-.08em
    T\kern-.1667em\lower.7ex\hbox{E}\kern-.125emX}}

\begin{document}

\title{EvoPath: An LLM-Agent Flow for False-Path Identification in Transistor-Level SRAM Timing Sign-Off\thanks{\hspace{-1em}\textsuperscript{*}This research was supported in part by a grant from Synopsys.}
}
\author{
\IEEEauthorblockN{Xuanyi Tan$^{1}$, Sanmitra Banerjee$^{1}$, Huiping Huang$^{2}$, Mayukh Bhattacharya$^{2}$ and Krishnendu Chakrabarty$^{1}$}
\IEEEauthorblockA{$^1$School of Electrical, Computer and Energy Engineering, Arizona State University, Tempe, AZ\\
% $^1$ASU Center for Semiconductor Microelectronics\\
$^2$Synopsys, Inc., Sunnyvale, CA\\
\vspace{-1.5em}
}
}

% \and
% \and
% \and
% \and
% \and

\maketitle
\thispagestyle{empty} 
% \vspace{-50em}\thispagestyle{plain} 

% \vspace{-100em}

\begin{abstract}
% Static random-access memory (SRAM) is critical to artificial intelligence (AI) accelerators, making efficient timing sign-off increasingly important. SRAM relies on transistor-level static timing analysis (STA), whose exhaustive path enumeration reports paths that are topologically feasible but functionally unsensitizable. Identifying these false paths requires design-specific functional knowledge and remains manual and expert-intensive. We present a gentic workflow that combines domain-specific agentic skills, graph-based circuit analysis, and progressive agentic memory to automate transistor-level false-path identification. The agent explores multiple graph views of the netlist to recover its topology and functionality, from which it constructs a functional plan with the tests and analog-node sensitization conditions required for path analysis. Paths are progressively screened for sensitizability through differential simulation, event-sound satisfiability (SAT) proofs, and SAT-guided SPICE simulation. Verified path knowledge is distilled into agentic memory, enabling the agent to reuse prior experience while exploring design-specific behaviors on unseen SRAMs. 
% Across 12 SRAMs, EvoPath agrees with switch-level symbolic analysis on all paths confined to digital logic and resolves the paths through analog nodes on which the switch-level abstraction diverges, while memory reuse reduces the effort on unseen designs.

Static random-access memory (SRAM) is critical to artificial intelligence accelerators, making efficient timing sign-off increasingly important. SRAM relies on transistor-level static timing analysis, which can report paths that are topologically feasible but functionally unsensitizable. Identifying these false paths requires design-specific functional knowledge and remains manual- and expert-intensive. We present EvoPath, an agentic workflow that combines domain-specific agentic skills, graph-based circuit analysis, and agentic memory to automate transistor-level false-path identification. The agent explores multiple graph views of the netlist to recover its topology and functionality, from which it constructs a functional plan with the tests and analog-node sensitization conditions required for path analysis. 
The agent recovers SRAM intent, while every timing exception is established by deterministic SAT or SPICE evidence.
Paths are screened for sensitizability through differential simulation, event-sound satisfiability (SAT) proofs, and SAT-guided SPICE validation. Verified path knowledge is distilled into agentic memory, enabling the agent to reuse experience while exploring design-specific behaviors on previously unseen SRAMs. 
Across 12 SRAM designs, EvoPath agrees with the switch-level symbolic reference in false-path identification and verification on digital paths, while complementing it on paths through analog nodes whose behavior is not represented by the switch-level abstraction. 

% Across 12 SRAMs, EvoPath remains faithful to transistor-level circuit behavior where a commercial symbolic simulation baseline misclassifies paths involving analog nodes, while memory reuse reduces the effort on unseen designs.
\end{abstract}
% \vspace{-0.6em}

\section{Introduction}
The rapid growth of artificial intelligence (AI) accelerators has heightened the
demand for dense, high-performance on-chip static random-access memory (SRAM)
\cite{sze2017efficient}, making timing sign-off important.
Unlike standard-cell logic, SRAM contains full-custom digital and analog
structures requiring transistor-level static timing analysis (STA).
Transistor-level STA conservatively enumerates feasible paths, including unsensitizable paths in the set of paths that are deemed to be sensitizable during functional operation~\cite{nanotime2025}. Identifying and excluding these false paths is essential to reduce timing pessimism but remains challenging to automate.
\par

This long-standing problem persists even though transistor-level STA has offered timing-exception mechanisms for decades~\cite{nanotime2025}. The intent needed to justify these exceptions still comes from experienced designers, while algorithmic approaches remain limited by their abstractions. Transistor-level timing studies focus on accuracy and scalability rather than false-path identification~\cite{kulshreshtha2000transistor,lee2018fast}. Gate-level false-path analysis assumes Boolean
gates that full-custom circuits lack~\cite{roy2008event}, while switch-level symbolic abstraction does not represent analog behaviors common in SRAMs~\cite{esp2026}. Manual practice remains
manageable for a small set of SRAM configurations, but growing custom
multi-port macros for AI accelerators and process, voltage, and temperature (PVT) sign-off make manual analysis a bottleneck.
Automation therefore requires recovering SRAM functional intent from the
netlist and verifying paths spanning digital and analog structures at the
transistor level.

% In practice, false-path identification requires design-specific knowledge, such as operating modes, analog partitions, and self-timed structures. Experienced designers typically provide this information because
% current STA methodologies cannot infer all required intent and analog constraints~\cite{nanotime2025}. Missing a
% false-path exception introduces timing pessimism, while an incorrect one may
% mask a real timing violation. 
% Prior transistor-level timing studies focus on accuracy and
% scalability, rather than
% false-path identification~\cite{kulshreshtha2000transistor,lee2018fast}. 
% Symbolic simulation tools such as ESP can analyze transistor-level paths, but their switch-level abstraction does not represent analog behaviors common in SRAMs~\cite{esp2026}. 
% False-path verification also requires path-specific functional constraints and verification setup. Addressing these gaps requires an automated approach that recovers SRAM functional intent and verifies paths spanning digital and analog structures at the transistor level.

% Formal verification tools such as ESP can analyze
% full-custom SRAMs, but false-path verification still requires path-specific
% functional constraints and verification setup~\cite{esp2026}. Constructing these verification problems requires substantial design knowledge and functional intent. To the best of our knowledge, automatic
% false-path identification remains an open challenge in
% transistor-level SRAM timing sign-off.

Agentic AI provides a promising way to automate this process. Recent large
language model (LLM)-based electronic design automation (EDA) systems show
that agents can decompose design tasks, invoke specialized tools, and refine
    outputs through tool feedback~\cite{wu2024chateda,qiu2024autobench,zhang2025llm4dv,tan2026safegen,
wang2026coverassert,lyu2026assertminer,he2025large, yan2026neuroabs, tan2025cotia}. LLM4SDC extends this paradigm to false-path discovery from specifications and register-transfer level (RTL) designs~\cite{han2026llm4sdc}, while AnalogTester and AnalogVerifier explore LLM-driven testbench generation and analog verification~\cite{chen2025analogtester,liuanalogverifier}. Agentic memory further enables
experience to be retained and reused across tasks without updating model
weights~\cite{zhao2024expel,wang2024agent,fang2026memp}. However, existing
approaches operate on higher-level design representations or
specialized analog tasks, leaving agentic transistor-level circuit
understanding and timing analysis largely unexplored. \par

% In this paper, we propose EvoPath, an agentic framework for false-path
% identification and verification in transistor-level SRAM timing sign-off. An
% LLM agent recovers SRAM topology and functionality from multi-view netlist
% graphs, constructs a functional plan, and proposes candidate paths. Every
% timing exception is then established by deterministic SAT or SPICE evidence.
% Differential SPICE first searches for sensitization witnesses, while an
% event-sound satisfiability (SAT) encoding proves unresolved paths false when
% UNSAT and seeds subsequent SPICE validation when SAT. The SAT proof is
% independent of PVT-dependent voltage and delay quantities. Verified results are
% distilled into progressive agentic memory that accelerates, but never decides,
% the analysis of unseen SRAMs.

In this paper, we propose EvoPath, an agentic framework for false-path identification and verification in transistor-level SRAM timing sign-off. An LLM agent recovers SRAM topology and functionality from multi-view graphs of the netlist and encodes them in a functional plan that drives the entire flow. Each STA-reported path is screened by differential SPICE simulation, then settled by an event-sound satisfiability (SAT) encoding: UNSAT proves the path false independently of delay and PVT corners, while SAT seeds subsequent SPICE validation. Verified results are distilled into a progressive agentic memory that accelerates, but never decides, the analysis of previously unseen SRAMs.

% we propose a self-evolving agentic framework for automatic
% false-path identification and verification in transistor-level SRAM timing
% sign-off. Domain-specific agentic skills orchestrate the flow, while multi-view graph
% analysis recovers circuit topology and functionality from the transistor-level
% netlist to generate functional SPICE tests and analog-node sensitization
% conditions. The framework supports an STA-assisted mode that analyzes STA-reported paths across process, voltage, and temperature (PVT) corners and an interactive mode for expert-guided path exploration. Functional tests first screen each path for sensitization
% through differential simulation. Unresolved paths then undergo event-sound satisfiability (SAT)
% sensitization analysis, where UNSAT proves the absence of propagating events
% independent of delay assignments, while SAT assignments seed paired SPICE
% simulations for full-path validation. Verified path knowledge is distilled
% into progressive agentic memory and reused across SRAMs. By balancing
% cross-design knowledge reuse with design-specific exploration, the framework
% continuously accumulates verified experience and accelerates analysis on
% unseen designs.

The key contributions of this paper are as follows:
\begin{itemize}
    \item We present, to the best of our knowledge, the first automated framework for
    false-path identification and verification in transistor-level SRAM
    timing sign-off.

    \item We introduce multi-view graph analysis that enables agents to
    understand SRAM topology and functionality from the 
    netlists and thereby automate false-path analysis.

    \item We develop a three-stage verification flow combining differential simulation, event-sound SAT proofs, and SAT-guided SPICE validation with characterized analog thresholds, verifying paths spanning digital and analog structures beyond switch-level abstraction.

    \item We propose a progressive agentic memory that distills verified knowledge across SRAMs, accelerating screening and exploration on unseen designs.
\end{itemize}

\section{Background}
\vspace{-0.4em}
\subsection{Transistor-Level SRAM Timing Analysis}
\vspace{-0.1em}
% As shown in Fig.~\ref{fig:ccb}(b), an SRAM macro integrates a bitcell array with full-custom peripheral and self-timed circuits ~\cite{guthaus2016openram,amrutur1998replica}. Due to its transistor-level implementation and analog behavior, SRAM timing sign-off is performed directly at the transistor level~\cite{nanotime2025,kulshreshtha2000transistor}.
% Transistor-level timing analysis operates on \emph{channel-connected blocks} (CCBs) rather than individual transistors. A CCB is a maximal group of transistors connected through source/drain terminals, while transistor gates
% bound the block~\cite{nanotime2025}. At the timing-graph level, a CCB serves as the basic timing unit, analogous to a standard cell in gate-level STA.
% As shown in Fig.~\ref{fig:ccb}(a), an inverter and the pass gate it drives belong to one CCB, while the following inverter starts another. CCB partitioning forms a directed graph, with edges connecting the channel outputs of one CCB to the gate inputs of another~\cite{kulshreshtha2000transistor}.

\begin{figure}[t]
    \centering
    \includegraphics[width=0.9\linewidth]{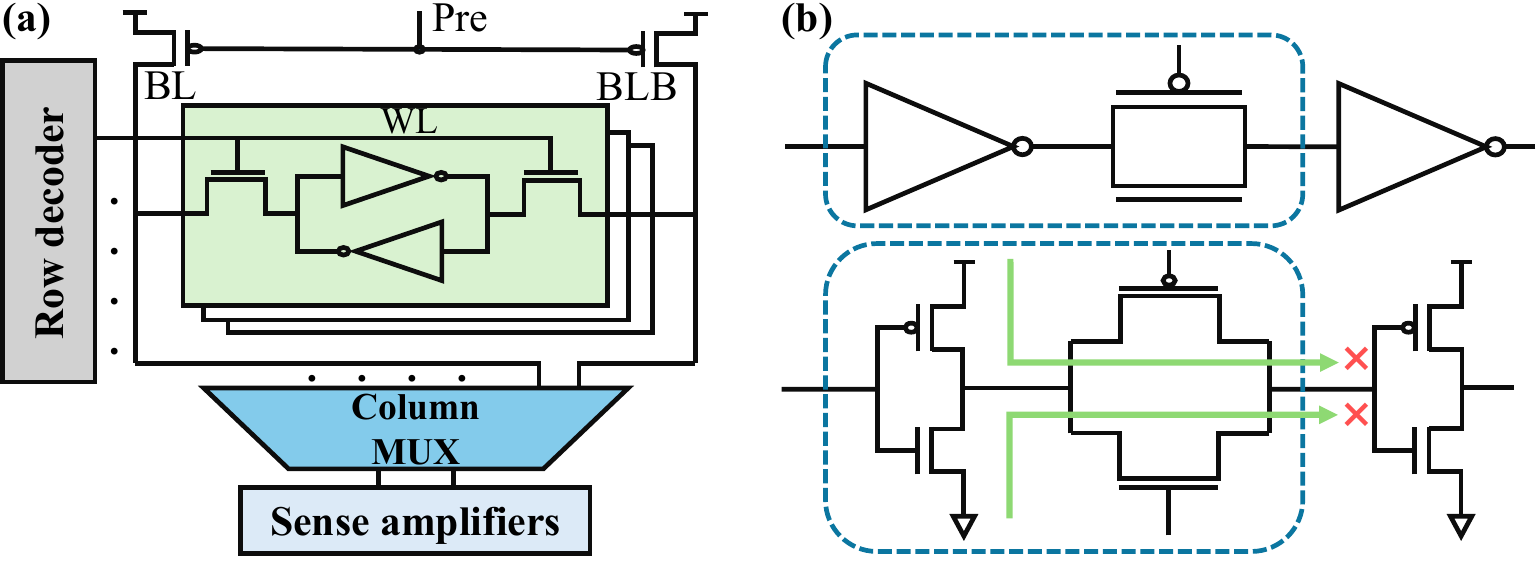}
    \vspace{-0.3em}
\caption{(a) SRAM macro organization. (b) Channel-connected block (CCB) partitioning of a transistor netlist.}
    \label{fig:ccb}
    \vspace{-0.8em}
\end{figure}

% Timing analysis then traces paths through the CCB graph. Transistor-level STA infers signal directions and circuit topology and characterizes timing arcs using local circuit simulation~\cite{nanotime2025}. Operating modes and timing exceptions further constrain the analysis. Since path tracing is conservative and
% vectorless, a topologically feasible path may not be sensitizable during functional operation. Such paths constitute the false-path problem discussed next.
As shown in Fig.~\ref{fig:ccb}(a), an SRAM macro integrates a bitcell array
with full-custom peripheral circuits~\cite{guthaus2016openram,amrutur1998replica}. Its transistor-level
implementation and analog behavior require timing sign-off on the transistor
network~\cite{nanotime2025,kulshreshtha2000transistor}.
To make timing analysis tractable, the circuit is partitioned into
channel-connected blocks (CCBs), which serve as basic timing units.
A CCB is a maximal group of transistors connected through source/drain
terminals and bounded by transistor gates~\cite{nanotime2025}. At the
timing-graph level, a CCB is analogous to a standard cell in
gate-level STA. As shown in Fig.~\ref{fig:ccb}(b), an inverter and the pass
gate it drives belong to one CCB, while the following inverter starts another.
The resulting timing graph connects the channel outputs of one CCB to the gate
inputs of another~\cite{kulshreshtha2000transistor}.

Transistor-level STA infers signal directions and traces paths through the CCB
graph, with timing arcs characterized by local circuit simulation~\cite{nanotime2025}. 
Operating modes and timing exceptions further constrain the analysis. Since path tracing is conservative and
vectorless, a topologically feasible path may not be sensitizable during functional operation. Such paths constitute the false-path problem discussed next.

% \begin{figure}[t]
%     \centering
%     \includegraphics[width=0.7\linewidth]{figure/sram.pdf}
%     % \vspace{-1em}
%     \caption{sram.}
%     \label{fig:sram}
%     \vspace{-1.5em}
% \end{figure}

\vspace{-0.4em}
\subsection{False-Path Identification and Verification}
\vspace{-0.1em}

A timing path is false if it cannot be sensitized under any functional
scenario~\cite{benkoski1990timing}.
Non-robust sensitization allows propagation under some delay assignments,
whereas robust sensitization requires propagation despite off-path delay
variations~\cite{benkoski1990timing,chen1993path}. Automatic test pattern generation (ATPG) and SAT have been
widely used to solve such sensitization problems~\cite{ashar1995functional,silva1998timing,e2002satisfiability, roy2008event, eggergluss2007sat, eggersgluss2010monsoon, grosse2009contradictory}.
Floating-mode sensitization assumes unknown internal states and requires the applied vector to establish path propagation~\cite{devadas1993computation}. In its Boolean SAT encoding, SAT provides a sensitizing vector and UNSAT proves unsensitizability~\cite{silva1998timing,e2002satisfiability,ringe1998path}.
Later work adds timing-aware, path-specific, compact, and slew-aware formulations~\cite{sauer2015phaeton,lai2017path,wu2020speeding,tsai2022slewfta}, but all rely on gate-level Boolean models. More recently, LLM4SDC derives false-path constraints from unexercised RTL branches~\cite{han2026llm4sdc}, but remains within digital logic, and an uncovered branch alone does not prove unsensitizability.
Switch-level symbolic simulation extends exhaustive analysis to transistor netlists, but resolving nodes to logic values does not directly represent analog behavior~\cite{esp2026, nopper2012symbolic}.
SAT-based SPICE equivalence~\cite{jha2025verisim} and analog-circuit reachability~\cite{dang2004verification} demonstrate formal reasoning at the transistor and analog levels, but false-path sensitization across mixed digital--analog transistor-level paths in SRAM designs remains unexplored.

% Existing methods therefore do not directly address
% full-custom SRAM paths spanning transistor-level digital logic, state
% elements, and analog nodes. 

% Later work further extends SAT-based sensitization with timing-aware and path-specific formulations ~\cite{sauer2015phaeton,lai2017path}. \par
% Existing false-path methods mainly assume gate-level or RTL representations. LLM4SDC~\cite{han2026llm4sdc}, for example, derives false-path constraints using RTL branch coverage, but unexercised logic does not prove unsensitizability.
% These methods do not directly apply to full-custom SRAMs, where sensitization must be analyzed at the transistor level.

% \subsection{Agentic Memory}
% Agentic memory enables LLM agents to retain experience across tasks
% without updating model weights. Prior work stores self-feedback, distilled
% experience, or reusable workflows for retrieval in subsequent
% tasks~\cite{shinn2023reflexion,zhao2024expel,wang2024agent,fang2026memp},
% reducing exploration.
% We adopt a progressive memory that distills verified false-path knowledge and
% updates it after each design. 
% Compact retrieval reduces token and exploration
% costs on new SRAMs. Since SRAMs share recurring structures,
% knowledge can be reused across designs without memory growing proportionally
% with the number of instances~\cite{guthaus2016openram}.

\section{Proposed Methodology}
\vspace{-0.1em}
% \subsection{Overview}
\begin{figure*}[t]
    \centering
    \includegraphics[width=0.75\linewidth]{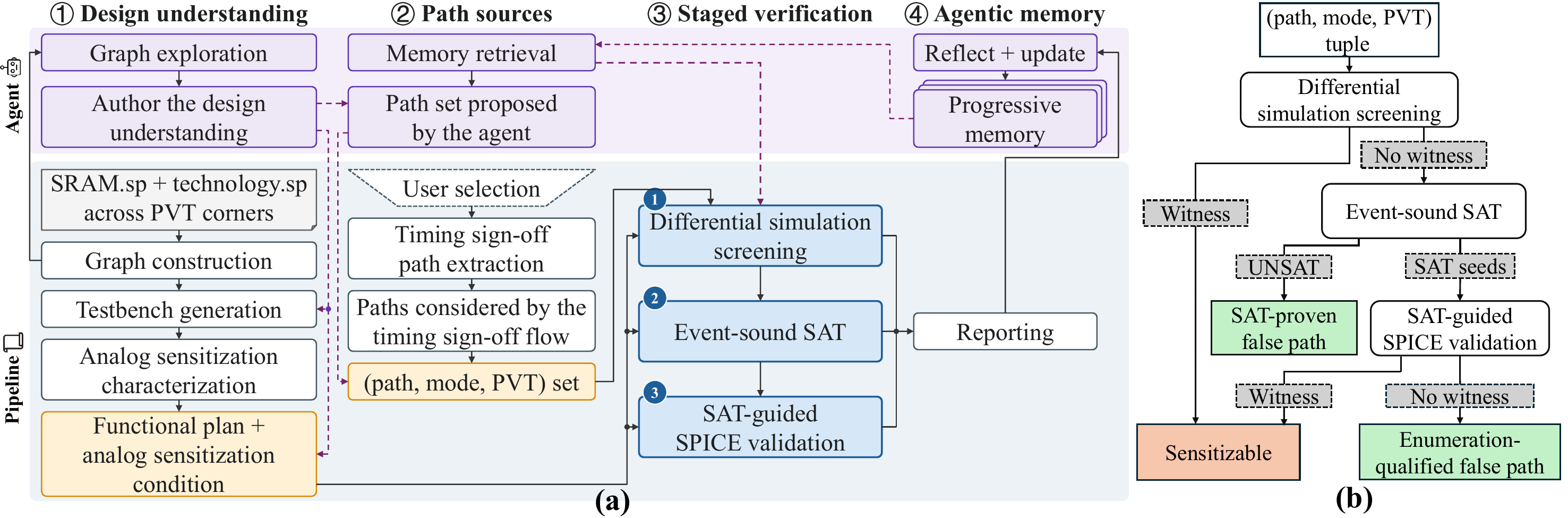}
    % \vspace{-1em}
\caption{(a) Skill-driven agentic framework for false-path identification and verification in SRAM timing sign-off. (b) Three-stage verification flow for each (path, mode, PVT) tuple.}
    \label{fig:flow}
    \vspace{-2em}
\end{figure*}

Fig.~\ref{fig:flow}(a) shows EvoPath, an agentic workflow for
false-path identification and verification in SRAM timing sign-off.

\vspace{-0.4em}
\subsection{Design Understanding}

\subsubsection{Multi-View Circuit Representation} 
% EvoPath parses the SPICE netlist using PySpice and constructs three
% cross-referenced graph views~\cite{pyspice}. The block graph folds repeated
% subcircuit instances into functional blocks connected by signal flow, providing
% a compact view of SRAM structures such as decoders, precharge circuits, sense
% amplifiers, and bitcell arrays. The hierarchy graph preserves the instance
% tree, with each instance annotated by its definition, port directions, and
% associated CCBs. The CCB graph captures transistor-level propagation by
% partitioning the circuit into connected CCBs. Together, these views represent
% the circuit at architectural, hierarchical, and transistor-connectivity levels.
% The agent directly accesses the compact block graph and queries the larger
% hierarchy and CCB graphs through bounded graph tools supporting hierarchical
% navigation, instance and CCB inspection, and forward/backward signal tracing.
EvoPath parses the SPICE netlist using PySpice and constructs three
cross-referenced graph views~\cite{pyspice}. The block graph groups repeated
subcircuits into functional blocks, providing a compact architectural view of
the SRAM. The hierarchy graph preserves the instance tree with definitions,
port directions, and associated CCBs, while the CCB graph represents
transistor-level signal propagation. Together, they capture architectural,
hierarchical, and transistor-connectivity information at complementary
abstraction levels. The agent directly accesses the block graph and queries
the larger hierarchy and CCB graphs through bounded tools for structural
inspection and signal tracing.

\subsubsection{Design Profiling and Testbench Generation}
\label{sec:tb}
Using the graph views, the agent records its design understanding in two
schema-constrained artifacts for subsequent analysis. 
The design
profile captures SRAM architecture, port roles, control relationships, and analog partitions, while the functional plan specifies timing parameters, functional SRAM operation modes, and legal
transitions. The agent enumerates all control combinations and assigns each to a functional
mode, with a deterministic coverage check ensuring that none is omitted. Simulation then verifies that the resulting plan reproduces the expected SRAM functionality.
Deterministic scripts render SPICE testbenches from the functional plan using
established SRAM workloads: complementary read/write patterns from March
SS~\cite{hamdioui2002march}, repeated accesses from SRAM Hammer
~\cite{hamdioui2006memory}, and repeated reads from read-destructive
tests~\cite{bosio2009advanced}. 
% These workloads cover diverse functional SRAM
% operations for analog characterization and sensitization
% screening in Section~\ref{sec:analog} and Section~\ref{sec:d}.
% EvoPath instantiates them with the modes, addresses, data, timing, and analog
% probes defined in the functional plan to generate legal SPICE stimuli for
% analog characterization and sensitization screening in
% Sections~\ref{sec:analog} and~\ref{sec:d}.

\subsubsection{Analog Sensitization Characterization} 
\label{sec:analog}

% EvoPath derives analog-node sensitization thresholds from circuit simulation, as bitline and sense-amplifier behavior depends on array loading and regeneration.
% At each PVT corner, EvoPath characterizes these analog nodes using the testbenches generated in~\cref{sec:tb}.
% The characterization collects two sets of observations: functional switching (e.g., during reads or writes) and non-switching activity caused by noise,
% coupling, and other circuit activity. 
% For each observation, EvoPath takes the node voltage immediately before the
% operation as the reference and measures the maximum voltage deviation from
% this value during the operation. For example, a bitline that starts at
% $1.0$\,V and discharges to $0.7$\,V during a read has a deviation of
% $0.3$\,V.
% For a given design, PVT corner, and analog-node class, let $\mathcal{S}$ and
% $\mathcal{N}$ denote the absolute voltage deviations, irrespective of
% polarity, of all functional-switching and non-switching observations,
% respectively. EvoPath defines $S_{\min}=\min\mathcal{S}$ and
% $N_{\max}=\max\mathcal{N}$, the weakest observed functional event and the
% strongest observed non-switching activity. Characterization is accepted only
% when $S_{\min}\geq 2N_{\max}$, and the sensitization threshold is set to the
% midpoint $\theta=(N_{\max}+S_{\min})/2$, so that $N_{\max}<\theta<S_{\min}$.
% % with a margin of at least $S_{\min}/4$ on either side.
% Section~\ref{sec:threshold_sensitivity} further evaluates the robustness of
% $\theta$ to threshold variation.
EvoPath derives analog-node sensitization thresholds from circuit simulation,
as bitline and sense-amplifier behavior depends on array loading and
regeneration. At each PVT corner, EvoPath characterizes these nodes using the
testbenches generated in~\cref{sec:tb}. The characterization collects
functional-switching observations (e.g., reads or writes) and non-switching
activity caused by noise, coupling, and other circuit activity.
For each observation, EvoPath takes the node voltage immediately before the
operation as the reference and measures the maximum voltage deviation from
this value during the operation.
For example, a bitline that
starts at $1.0$\,V and discharges to $0.7$\,V during a read has a deviation of
$0.3$\,V. For a given design, PVT corner, and analog-node class, let
$\mathcal{S}$ and $\mathcal{N}$ denote the sets of absolute deviations of all
functional-switching and non-switching observations, respectively. EvoPath
defines $S_{\min}=\min\mathcal{S}$ and $N_{\max}=\max\mathcal{N}$ as the
weakest functional event and strongest non-switching activity. Characterization
is accepted only when $S_{\min}\geq2N_{\max}$, and the sensitization threshold
is $\theta=(N_{\max}+S_{\min})/2$. Section~\ref{sec:threshold_sensitivity}
evaluates its robustness to threshold variation.
% At each PVT corner, EvoPath characterizes representative bitline and
% sense-amplifier nodes using the testbenches generated
% in~\cref{sec:tb}. The characterization observes each analog node under two conditions: functional switching during read or write operations, and non-switching behavior that captures noise, coupling, and other background activity.
% For each observation, EvoPath measures the largest voltage deviation from the
% node value immediately before the operation. For example, if a bitline starts
% at $1.0$\,V and drops to $0.7$\,V during a read, its measured deviation is
% $0.3$\,V. The smallest deviation among valid events is denoted by
% $S_{\min}$, while the largest deviation among background observations is
% denoted by $N_{\max}$. EvoPath accepts the characterization only when
% $S_{\min}\geq2N_{\max}$ and sets the sensitization threshold to
% $\theta=\sqrt{N_{\max}S_{\min}}$. By construction,
% $N_{\max}<\theta<S_{\min}$, so the threshold lies above all observed
% background activity and below all observed valid events. We further evaluate
% its robustness through threshold-sensitivity analysis in
% Section~\ref{sec:threshold_sensitivity}.

\vspace{-0.4em}
\subsection{Path Sources}
\vspace{-0.2em}
EvoPath acquires paths in two ways. The STA-assisted mode extracts CCB-based
paths reported by transistor-level STA, while
the interactive mode allows users to explore and identify
paths of interest with the agent. Each path is evaluated under a functional
SRAM mode defined in the functional plan, such as read, write, or idle, and at a
PVT corner, forming a (path, mode, PVT) tuple for verification in
Section~\ref{sec:sim}.
\vspace{-0.25em}
\vspace{-0.4em}
\subsection{Staged Verification}
\label{sec:sim}
\vspace{-0.2em}
EvoPath verifies each (path, mode, PVT) tuple through three stages, as shown
in Fig.~\ref{fig:flow}(b). Differential SPICE simulation searches for a
sensitization witness~\cite{sauer2015phaeton}, followed by event-sound SAT analysis for unresolved
cases and SAT-guided SPICE validation. Both simulation stages evaluate the
complete CCB-level path, including analog nodes, whereas SAT operates on its
Boolean prefix. The Boolean prefix is the path segment from the source to the
first CCB that cannot be represented by a Boolean function, such as an analog structure.
% EvoPath extracts its fan-in cone for SAT analysis. 
% Since any event
% traversing the complete path must first propagate through this prefix, proving
% the prefix unsensitizable is sufficient to prove the corresponding
% (path, mode, PVT) tuple false.

\subsubsection{Sensitization Criterion}
\label{sec:sense}

% For both SPICE simulation stages, EvoPath defines path sensitization as
% time-aware event propagation from the source to the endpoint. 
% The STA report provides the expected rise or fall at each
% path node. For digital nodes, a valid event must exceed the $V_{\mathrm{DD}}/2$ with the expected polarity and arrive in path order. Transient pulses that
% satisfy these conditions are also treated as events.
% Analog nodes use the corner-specific thresholds derived in
% Section~\ref{sec:analog}. Their internal arrival order is not enforced because
% bitline and sensing activity can begin before subsequent control signals
% switch, making threshold-crossing order an unreliable indicator of propagation
% order. The endpoint must occur after the preceding path events to
% establish that the event reaches the end of the path. A path is sensitized
% only when the required events propagate from the source through the endpoint.

For both SPICE simulation stages, EvoPath defines path sensitization as time-aware event propagation from the source to the endpoint. The STA report provides the expected rise or fall at each path node. For digital nodes, a valid event must exceed $V_{\mathrm{DD}}/2$ with the expected polarity and arrive in path order. 
% Transient pulses satisfying these conditions also count as events. 
Analog nodes use the corner-specific thresholds derived in Section~\ref{sec:analog}. Their arrival order is not enforced because bitline and sensing activity can begin before subsequent control signals switch, making threshold-crossing order an unreliable indicator of propagation order. The endpoint event must occur after the preceding path events to establish that the event reaches the end of the path. A path is sensitized only when the required events propagate from the source through the endpoint.

\subsubsection{Differential Simulation Screening}
\label{sec:d}
% EvoPath first screens each (path, mode) pair using the functional SPICE testbenches generated in ~\cref{sec:tb}. Only scenarios consistent with the corresponding operating mode are used, so sensitization is evaluated under
% the functional constraints of that mode. Each test is simulated twice: a launch run containing the source event and a matched control run with only that event suppressed. For an ordered path $P=(n_0,\ldots,n_K)$, let $V_i^L(t)$ and $V_i^C(t)$ denote the waveforms at node $n_i$ from the launch and control runs, respectively. Their difference is $\Delta_i(t)=V_i^L(t)-V_i^C(t)$,
% which isolates waveform changes caused by the source event from normal SRAM activity shared by both runs.
% Within the observation window, EvoPath applies the sensitization criteria defined above to $\Delta_i(t)$. The differential must exhibit the expected rise or fall at every path node; digital events must propagate in path order,
% while analog nodes use their characterized thresholds without the ordering requirement. A transient pulse is retained as an event if it crosses the corresponding threshold within the window. If these conditions hold through
% the endpoint, the simulation provides a concrete sensitization witness and the \emph{(path, mode)} pair is classified as sensitizable. Otherwise, the pair remains unresolved and proceeds to SAT analysis.
Differential simulation first searches for sensitization witnesses using the
workloads in Section~\ref{sec:tb}. In each (path, mode, PVT) tuple,
mode specifies the SRAM mode in which the path is tested, while the preceding
operation may use any mode allowed by the functional plan. Thus, a path tested in read mode covers transitions such as read-to-read, idle-to-read, and write-to-read, each with address and data variations.

For each sensitization check, EvoPath performs two matched SPICE simulations. The launch simulation
contains the STA-reported source transition during the tested operation,
whereas the control simulation suppresses only this transition. For node
$n_i$, let $V_i^L(t)$ and $V_i^C(t)$ denote its launch and control waveforms.
EvoPath computes $\Delta_i(t)=V_i^L(t)-V_i^C(t)$ and applies the sensitization
criteria in Section~\ref{sec:sense}, isolating the response caused by the
source transition. If the event reaches the endpoint, the simulation provides
a sensitization witness. Otherwise, the tuple proceeds to SAT analysis, since
this stage only screens for witnesses and cannot establish a false path.

\subsubsection{Event-Sound SAT Sensitization Analysis}
\label{sec:sat}
% For each case not witnessed by differential simulation, EvoPath asks whether
% a legal predecessor operation followed by the current-mode operation can
% propagate an event through the Boolean prefix. The encoding follows a strict
% rule: every constraint must be necessary for a physical sensitization witness;
% when circuit behavior cannot be safely modeled, EvoPath omits constraints
% rather than introducing assumptions. Hence, UNSAT proves a false path, whereas
% SAT only identifies a condition for subsequent SPICE validation. Since the
% encoding contains no PVT-dependent voltage or delay quantities, the UNSAT
% proof is independent of the PVT corner.
For each case unresolved by differential simulation, EvoPath uses SAT to
determine whether any legal mode transition can propagate an event through the
Boolean prefix. The encoding includes only conditions necessary for a physical
sensitization witness and leaves unmodeled behavior unconstrained. UNSAT
therefore proves a false path independently of PVT corner, while SAT provides a condition for subsequent SPICE validation.

\textbf{Settled Frames and Boolean Model.}
Consider a path $\pi=(n_0,n_1,\ldots,n_K)$ with Boolean prefix
$\pi_B=(n_0,n_1,\ldots,n_B)$. The STA report specifies a rise or fall at
the source node $n_0$, the only external stimulus in the analysis window.
EvoPath represents the window with two settled frames: $C$, the steady
state before the source event, and $L$, the steady state after it. From
$\pi_B$ and its fan-in cone, which captures the on-path logic and the side
inputs that determine event propagation, EvoPath constructs a SAT model in
which every Boolean relation of the cone is instantiated in both frames.

\begin{figure}[t]
    \centering
    \includegraphics[width=1\linewidth]{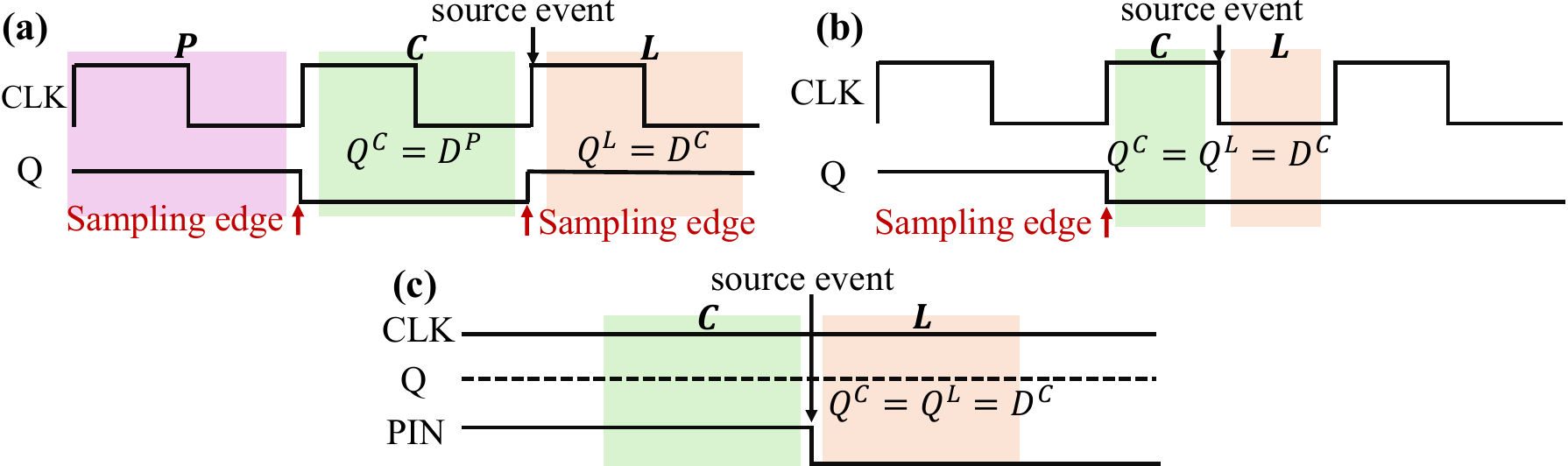}
    \caption{Register modeling across settled frames. (a) Sampling-edge launch: the edge falls between $C$ and $L$, and the value
in $C$ comes from the preceding operation in frame $P$. (b) Opposite-edge
launch and (c) input-pin launch: no sampling edge lies within the window,
so $Q$ holds its latched value.
    }
    \label{fig:regframes}
    \vspace{-0.5em}
\end{figure}

A CCB is represented by a Boolean function only when its transistor connectivity defines a unique static output for every input assignment. The resulting function
$y=f(x_1,\ldots,x_k)$ is encoded in conjunctive normal form (CNF). CCBs
with contention, floating behavior, feedback, or excessive input width are
not Booleanized. Their output $z$ remains a SAT variable without defining
clauses, allowing both $z=0$ and $z=1$, while cone extraction continues
through the CCB inputs to preserve the upstream Boolean logic.
% Registers are modeled separately according to their capture behavior, which
% depends on the path source. For a path launched from the capture (rising)
% edge of the clock, the register state before the edge was established by
% the preceding operation, which may run in a different mode: a read-mode
% path may follow a write, for example, so the pre-edge state must reflect
% the write rather than the read-mode controls. EvoPath therefore adds a
% predecessor frame $P$, which may take any legal predecessor mode, and
% enforces $Q^C=D^P$ and $Q^L=D^C$ for a register with input $D$ and output
% $Q$, i.e., the values captured by the preceding and current operations,
% respectively. For a path launched from the falling edge of the clock, the
% launch cycle's own rising edge has already captured the current pins, so
% the register holds $Q^C=Q^L=D^C$ throughout the window. For a path launched
% from an input pin, $Q^C=D^C$ and $Q^L=D^L$.
All registers in the fan-in cone are modeled separately. Since a register
latches $D$ only at its sampling edge and holds $Q$ otherwise, the modeling
depends on whether the source event itself is a sampling edge. For a clock-launched path on the sampling edge
(Fig.~\ref{fig:regframes}(a)), the edge lies between the frames and
$Q^L=D^C$, whereas $Q^C$ is latched by the preceding operation, which may
run in a different mode (e.g., a write before a read path). EvoPath
therefore adds a predecessor frame $P$, constrained only to be a legal
mode, and sets $Q^C=D^P$. 
When no sampling edge occurs within the analysis window, including
opposite-edge clock-launched paths (Fig.~\ref{fig:regframes}(b)) and
input-pin-launched paths (Fig.~\ref{fig:regframes}(c)), each register holds
its latched value, $Q^C=Q^L$. 
Specifically, $Q^C=Q^L=D^C$ if the register's
fan-in cone contains only inputs held fixed across the preceding sampling edge
by the functional plan, and $Q^C=Q^L=D^P$ otherwise.

% In the designs considered here
% the sampling edge is the rising edge, so the two clock-launched cases
% correspond to rising- and falling-edge launches.

% For the Boolean prefix $\pi_B=(n_0,\ldots,n_B)$, EvoPath constrains the source
% $n_0$ to its STA-reported transition and fixes the control inputs according to
% the functional mode being analyzed. Inputs permitted to change by that mode
% remain unconstrained, while register crossings follow the frame relations
% defined above. If any on-path node is forced to a constant by these
% constraints, the source event cannot propagate through $\pi_B$ and the
% formula is UNSAT.

% For each Boolean CCB on $\pi_B$, EvoPath asks whether it can guarantee blocking
% all possible events. Let $v=f(a,\mathbf{s},\mathbf{u})$, where $a$ is the
% on-path input, $\mathbf{s}$ denotes side inputs proven stable, and
% $\mathbf{u}$ denotes inputs that may change. For
% $\mathbf{s}=\boldsymbol{\sigma}$, the CCB guarantees blocking only if
% $f(a,\boldsymbol{\sigma},\mathbf{u})=c_{\boldsymbol{\sigma}}$ for all $a$ and
% $\mathbf{u}$, i.e., its output remains constant regardless of any input that
% may change. EvoPath then adds a clause excluding
% $\boldsymbol{\sigma}$; otherwise, the assignment remains feasible because
% some relative timing may still produce an output event. As shown in
% Figure.~\ref{fig:sat}, this rule preserves transient pulses and events entering
% through reconvergent side inputs.

\begin{figure}[t]
    \centering
    \includegraphics[width=1\linewidth]{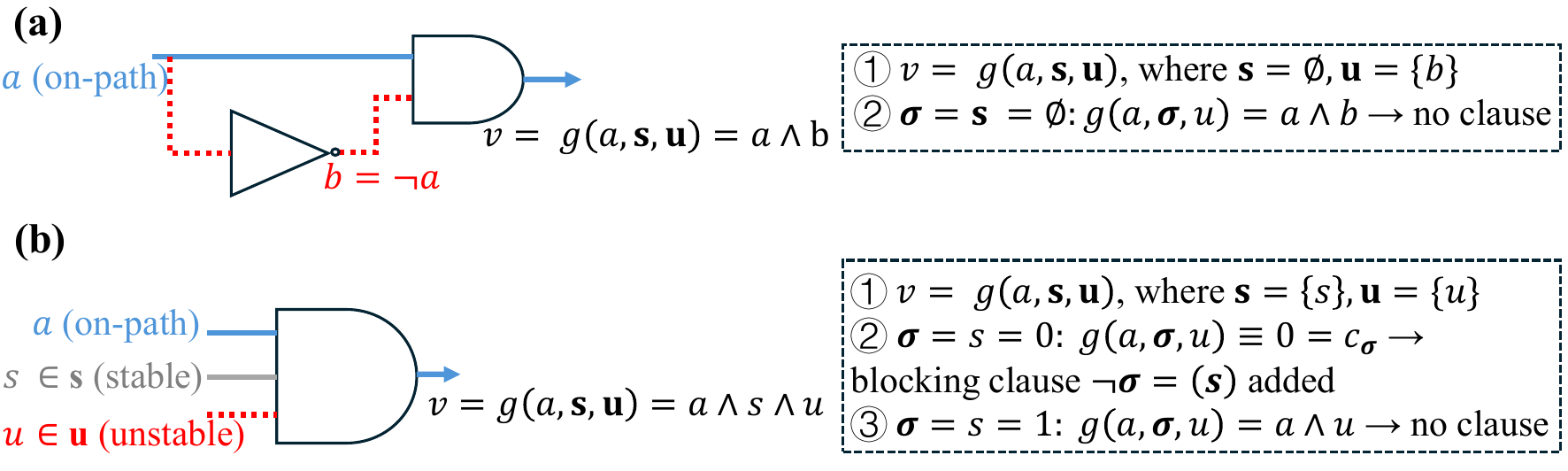}
    % \vspace{-1em}
\caption{Event-blocking examples: (a) an unstable side input preserves a
possible transient event, so no blocking clause is added; (b) a stable side
input is enumerated, with $s{=}0$ excluded by the blocking clause $(s)$.}
    \label{fig:sat}
    \vspace{-0.7em}
\end{figure}

\textbf{Encoding.}
On top of the Boolean model, EvoPath adds a source, a mode, and a blocking
constraint.
The source constraint assigns $n_0^C$ and $n_0^L$ the values before and
after the STA-reported transition.
The mode constraint reflects the operation being analyzed, such as a read
or a write. For each operation, the functional plan specifies which
control pins are held fixed, which pins may change, and which operations
may legally precede it. An input is stable if the operation or
differential simulation holds it constant across $C$ and $L$, and unstable
otherwise. Stable inputs are fixed in both frames, unstable inputs are
left free since they may switch at any time relative to the source event,
the legal predecessors bound frame $P$, and registers follow the frame
relations defined above.
The blocking constraint asks, for each Boolean CCB on $\pi_B$, whether it
can guarantee blocking all events. Let the CCB output be
$v=g(a,\mathbf{s},\mathbf{u})$, where $g$ is its Booleanized function with
inputs partitioned into the on-path input $a$, stable side inputs
$\mathbf{s}$, and unstable side inputs $\mathbf{u}$. For an assignment
$\boldsymbol{\sigma}$ of $\mathbf{s}$, the CCB blocks only if
$g(a,\boldsymbol{\sigma},\mathbf{u})=c_{\boldsymbol{\sigma}}$ for all $a$
and $\mathbf{u}$, where $c_{\boldsymbol{\sigma}}$ is the resulting
constant output. EvoPath then adds a clause excluding $\boldsymbol{\sigma}$.
Otherwise the assignment remains feasible, since
some relative timing of the unstable inputs may still produce an output
event. As shown in Fig.~\ref{fig:sat}, this preserves transient pulses and
events entering through reconvergent side inputs.

We next present a lemma and proof based on the following assumptions: (A1) Booleanized CCBs follow their extracted functions,
(A2) registers follow their sampling behavior, (A3) the functional plan specifies the operations and all their legal predecessors, and (A4) an unstable input may change at any time relative
to the source event.
\vspace{-0.7em}
\begin{lemma}[UNSAT soundness]
Under A1--A4, if the Boolean-prefix encoding $\Phi(\pi_B,m)$ for operation
$m$ is UNSAT, the complete path cannot be sensitized under $m$.
\end{lemma}
\vspace{-0.6em}

\vspace{-0.7em}
\begin{proof}
Let $w$ be a witness: a physical execution, at any PVT corner, in which
the source event propagates through $\pi_B$ under~$m$. Reading the
settled values of $w$ in frames $P$, $C$, and $L$ gives an assignment
$\alpha_w$. By A1 and A2, $\alpha_w$ satisfies the Boolean model. By A3
and the reported transition, it satisfies the mode and source constraints.
Non-Booleanized CCBs add no clauses. 
A blocking clause forbids a stable
side-input value $\boldsymbol{\sigma}$ only if, under
$\boldsymbol{\sigma}$, the CCB output is constant for all on-path and
unstable-input values. Since by A4 an unstable input may switch at any
time, this covers every possible timing, leaving the output unchanged
across frames so that no event can propagate through the CCB. Since the
event does pass every CCB in~$w$, $\alpha_w$ avoids every forbidden
$\boldsymbol{\sigma}$. Thus every witness satisfies $\Phi(\pi_B,m)$, UNSAT means that no witness exists for $\pi_B$, and hence none for the
complete path, which must traverse~$\pi_B$.
\end{proof}
\vspace{-0.7em}

Because the encoding lacks voltage or delay quantities, UNSAT proves
a false path independently of the PVT corner.

\subsubsection{SAT-Guided SPICE Validation}
\label{sec:sat_spice}
% A satisfiable encoding yields candidate sensitization conditions that must
% be confirmed by simulation, but its satisfying assignments are numerous
% and many differ only in internal Boolean variables while implying the same
% stimulus. EvoPath projects each assignment onto the controllable
% variables, namely the input values, bus transitions, and the predecessor
% operation that establishes the required register states, to form a
% simulation seed.

% A concrete SRAM address is also derived for each seed to exercise the corresponding
% path. EvoPath takes the last Boolean node of the path and its STA-reported
% transition as the target. Since this node is a Boolean function of its fan-in,
% fixing the mode controls and the target value allows SAT to solve backward for
% an address that produces the required response. For example, constraining a
% wordline to its active value solves the decoder backward from the wordline to
% the address inputs. The seed and derived address are then used to generate the
% SPICE testbenches.

% Each seed is evaluated by differential simulation. If the event propagates, the (path, mode, PVT) tuple
% is established as sensitizable. Otherwise, a blocking clause over the
% projected variables excludes the seed and SAT is solved again, so that
% each iteration yields a distinct stimulus. The loop ends when a witness is found or
% the seeds are exhausted. Unlike an UNSAT proof, exhaustion is specific to
% the simulated PVT corner and to the enumerated seeds, and EvoPath reports
% it as an enumeration-qualified false path for that corner.
A satisfiable encoding yields candidate sensitization conditions, but many
satisfying assignments differ only in internal variables and imply the same
stimulus. EvoPath projects each assignment onto the controllable variables,
namely the input values, bus transitions, and the predecessor operation, to
form a simulation seed. A concrete SRAM address is derived alongside each
seed: EvoPath fixes the mode controls and the STA-reported transition at the
last Boolean node, then solves SAT backward through the fan-in logic to obtain
a matching address. For example, constraining a wordline to its active value
solves the decoder backward to the address inputs.

Each seed is evaluated by differential simulation. If the event propagates, the
(path, mode, PVT) tuple is sensitizable. Otherwise, a blocking clause excludes
the seed's projected variables and SAT is re-solved, ensuring a distinct stimulus
at each iteration. The loop terminates when a witness is found or all seeds are
exhausted. Unlike an UNSAT proof, exhaustion is corner- and seed-specific, so
EvoPath reports the path as an enumeration-qualified false path for that corner.
\vspace{-0.4em}
\subsection{Progressive Agentic Memory}
\label{sec:memory}
\vspace{-0.1em}
Most SRAMs share recurring circuit structures, making verified false-path
experience reusable across designs. EvoPath organizes verified paths into
structural families defined by the source transition, CCB-role
sequence, and endpoint transition. These functional roles make the family
representation independent of design-specific net and instance names.
For a path matching a known family, EvoPath prioritizes the workloads and
stimuli that produced witnesses in previous designs during differential
simulation screening. If any stimulus sensitizes the path,
screening terminates early. Otherwise, the path proceeds through the full
workload set as in the standard flow.

In interactive mode, the family key provides direct access to relevant
experience, allowing the agent to retrieve matched entries rather than
revisit the full history. The retrieved structural knowledge is mapped to the
new SRAM to guide candidate-path exploration, reducing exploration effort
and agent tokens. Unseen structures are explored separately, and all candidates
undergo the same verification flow. Memory never determines a path
result: it accelerates differential simulation screening and exploration,
while unresolved paths still require SAT analysis and SAT-guided SPICE
validation.

\vspace{-0.2em}
\section{Experimental Evaluation}

\vspace{-0.1em}
\subsection{Experimental Setup}
\vspace{-0.1em}
All experiments are conducted on a 32-thread AMD EPYC 7313 workstation.
We use Claude Code with Claude Opus 5 as the agent harness, Synopsys
NanoTime for transistor-level STA, and Synopsys ESP as the switch-level
symbolic reference~\cite{nanotime2025,esp2026}.
We evaluate 12 OpenRAM-generated SRAMs~\cite{guthaus2016openram} in the
Nangate 45\,nm PDK with diverse configurations as shown in
Table~\ref{tab:designs}. Designs d01--d08 are analyzed without agentic memory
to accumulate verified experience, which is then evaluated on d09--d12 for
memory reuse.

\vspace{-0.4em}
\subsection{Robustness of Analog Sensitization Thresholds}
\label{sec:threshold_sensitivity}
\vspace{-0.1em}
To evaluate the robustness of the characterized analog sensitization threshold
$\theta$, we sweep the threshold from $N_{\max}$ to $S_{\min}$ in 20\,mV
steps and re-evaluate all paths involving analog nodes.
Table~\ref{tab:robust} reports the results for sense-amplifier nodes. Their
measured non-switching activity is below $10^{-4}$\,V across all tested cases,
so $N_{\max}$ is effectively zero. The stable interval in the table denotes
the threshold range over which all detected transitions and sensitization
results remain identical. Across all designs and PVT corners, the characterized
$\theta$ lies well inside the stable interval, with substantial margins to both
bounds. Bitline nodes are omitted because their sensitization results remain
unchanged for all tested thresholds up to $S_{\min}$.

% These results show that
% the characterized thresholds have substantial margin against threshold
% variation.

\vspace{-0.4em}
\subsection{False-Path Identification}

\vspace{-0.1em}

\begin{table}[t]
\centering
\caption{SRAM designs used in the evaluation.}
\vspace{-0.5em}
\label{tab:designs}
\scriptsize
\setlength{\tabcolsep}{5pt}            % column spacing
\renewcommand{\arraystretch}{0.05}      % row spacing (1.0 = default)
\begin{tabular}{@{}l l r c c c@{}}
\toprule
Set & Design & Capacity & Words $\times$ bits & Ports & \#\,Column MUX \\
\midrule
\multirow{8}{*}{\shortstack[l]{False-path\\evaluation}}
 & d01 & 8\,KB   & $2048\times32$   & 1RW    & 4  \\
 & d02 & 4\,KB   & $4096\times8$    & 1RW    & 16 \\
 & d03 & 16\,KB  & $1024\times128$  & 1RW    & 4  \\
 & d04 & 16\,KB  & $8192\times16$   & 1RW    & 16 \\
 & d05 & 4\,KB   & $1024\times32$   & 2RW    & 4  \\
 & d06 & 16\,KB  & $2048\times64$   & 1W+1R  & 8  \\
 & d07 & 8\,KB   & $512\times128$   & 1RW+1R & 1  \\
 & d08 & 8\,KB   & $512\times128$   & 1W+1R  & 2  \\
\midrule
\multirow{4}{*}{\shortstack[l]{Memory\\reuse}}
 & d09 & 128\,KB & $16384\times64$  & 1RW    & 8  \\
 & d10 & 64\,KB  & $16384\times32$  & 1RW+1R & 8  \\
 & d11 & 64\,KB  & $2048\times256$  & 1RW    & 2  \\
 & d12 & 128\,KB & $32768\times32$  & 1RW    & 16 \\
\bottomrule
\end{tabular}
\vspace{-1em}
\end{table}

\subsubsection{Switch-Level Symbolic Reference}
\label{sec:esp}
Since no automated transistor-level false-path tool exists, we compare
EvoPath's classifications of all paths against ESP~\cite{esp2026}, a switch-level symbolic simulator. ESP models devices as switches
with characterized RC delays and checks sensitization over multiple Boolean
assignments simultaneously without vector enumeration. Agreement on digital
paths cross-validates EvoPath, whereas divergence on analog-involving paths
locates where switch-level abstraction departs from transistor-level
simulation.
For a matched setting, we characterize ESP transistor models for each SRAM's
device sizes and PVT corners, configure floating-node and hierarchical
behavior, and calibrate switch-level RC delays against SPICE. Each path is
evaluated with matched launch and control copies under the same mode
transitions as EvoPath. ESP checks whether the differential event propagates
with the required polarity and ordering.

\subsubsection{Identification Results}
\begin{table}[t]
\centering
\caption{Threshold robustness for sense-amplifier nodes. Tuples report values at (SS, TT, FF), respectively.}
\vspace{-0.5em}
\label{tab:robust}
\scriptsize
\setlength{\tabcolsep}{1pt}
\renewcommand{\arraystretch}{1.1}
\resizebox{\columnwidth}{!}{%
\begin{tabular}{|c|c|c|c|}
\hline
Design & $S_{\min}$ (V) & $\theta$ (V) & Stable interval (V) \\
\hline
d01 & (0.563, 0.582, 0.602) & (0.282, 0.291, 0.301) & ($[0.12,\,0.56]$, $[0.16,\,0.58]$, $[0.22,\,0.60]$) \\
d02 & (0.542, 0.554, 0.577) & (0.271, 0.277, 0.289) & ($[0.12,\,0.54]$, $[0.16,\,0.54]$, $[0.20,\,0.56]$) \\
d03 & (0.563, 0.582, 0.602) & (0.282, 0.291, 0.301) & ($[0.04,\,0.56]$, $[0.14,\,0.58]$, $[0.18,\,0.60]$) \\
d04 & (0.504, 0.522, 0.521) & (0.252, 0.261, 0.260) & ($[0.12,\,0.50]$, $[0.16,\,0.52]$, $[0.20,\,0.52]$) \\
d05 & (0.393, 0.409, 0.424) & (0.196, 0.205, 0.212) & ($[0.04,\,0.38]$, $[0.02,\,0.40]$, $[0.04,\,0.42]$) \\
d06 & (0.545, 0.530, 0.583) & (0.273, 0.265, 0.291) & ($[0.08,\,0.54]$, $[0.04,\,0.52]$, $[0.02,\,0.58]$) \\
d07 & (0.553, 0.572, 0.592) & (0.277, 0.286, 0.296) & ($[0.12,\,0.54]$, $[0.12,\,0.56]$, $[0.12,\,0.58]$) \\
d08 & (0.563, 0.582, 0.602) & (0.282, 0.291, 0.301) & ($[0.02,\,0.56]$, $[0.02,\,0.58]$, $[0.02,\,0.60]$) \\
\hline
\end{tabular}%                       
}
\vspace{-1.2em}
\end{table}

\begin{table*}[t]
\centering
\caption{False-path identification results reported by (path, mode) pair. All classifications are corner-independent across three evaluated PVT corners. Runtime is averaged across corners. A pair is digital-only when every node on its path is a digital node. The remaining pairs cross at least one analog node.}
\vspace{-0.5em}
\label{tab:false_path_results}
\scriptsize
\setlength{\tabcolsep}{5pt}
\renewcommand{\arraystretch}{0.7}
\resizebox{\textwidth}{!}{%
\begin{tabular}{|c|cccc|cccc|cc|cc|}
\hline
\multirow{2}{*}[-4pt]{Design}
& \multicolumn{4}{c|}{NanoTime}
& \multicolumn{4}{c|}{EvoPath}
& \multicolumn{2}{c|}{ESP}
& \multicolumn{2}{c|}{Disagreement (\#\,pairs)} \\
\cline{2-13}
& \makecell{\#\,Extracted} & \makecell{\#\,Digital-only\\pairs} & \makecell{\#\,Pairs with\\analog nodes} & \makecell{Runtime\\(h)}
& \makecell{\#\,Sensitizable} & \makecell{\#\,SAT-proven\\false} & \makecell{\#\,Enumeration-\\qualified false} & \makecell{Runtime\\(h)}
& \makecell{\#\,False} & \makecell{Runtime\\(h)}
& \makecell{EvoPath false\\ESP sensitizable} & \makecell{ESP false\\EvoPath sensitizable} \\
\hline
d01 & 16599 & 16191 & 408   & 0.30 & 14727 & 1803 & 69 & 1.38 & 1936 & 0.21 & 0 & 64 \\
d02 & 8217  & 8097  & 120   & 0.14 & 7353  & 843  & 21  & 0.52 & 880  & 0.10 & 0 & 16 \\
d03 & 80931 & 79371 & 1560  & 1.99 & 67342 & 13325 & 264 & 1.59 & 13845 & 0.77 & 0 & 256 \\
d04 & 80631 & 80451 & 180   & 3.16 & 68197 & 12403 & 31  & 2.24 & 12460 & 0.21 & 0 & 26 \\
d05 & 81414  & 58374  & 23040 & 31.29 & 58368  & 12768 & 10278 & 19.70 & 19590 & 64.00  & 3840 & 384 \\
d06 & 219468 & 189772 & 29696 & 4.85  & 125422 & 85538 & 8508  & 36.67 & 92358 & 162.50 & 2066 & 378 \\
d07 & 81288  & 67272  & 14016 & 1.57  & 59820  & 11845 & 9623  & 23.23 & 21468 & 1.73  & 0    & 0   \\
d08 & 46668  & 30284  & 16384 & 2.68  & 30246  & 10754 & 5668  & 7.91  & 14886 & 0.75  & 2048 & 512 \\
\hline
\end{tabular}%
}
\vspace{-2.2em}
\end{table*}

\begin{figure}[t]
    \centering
    \includegraphics[width=1\linewidth]{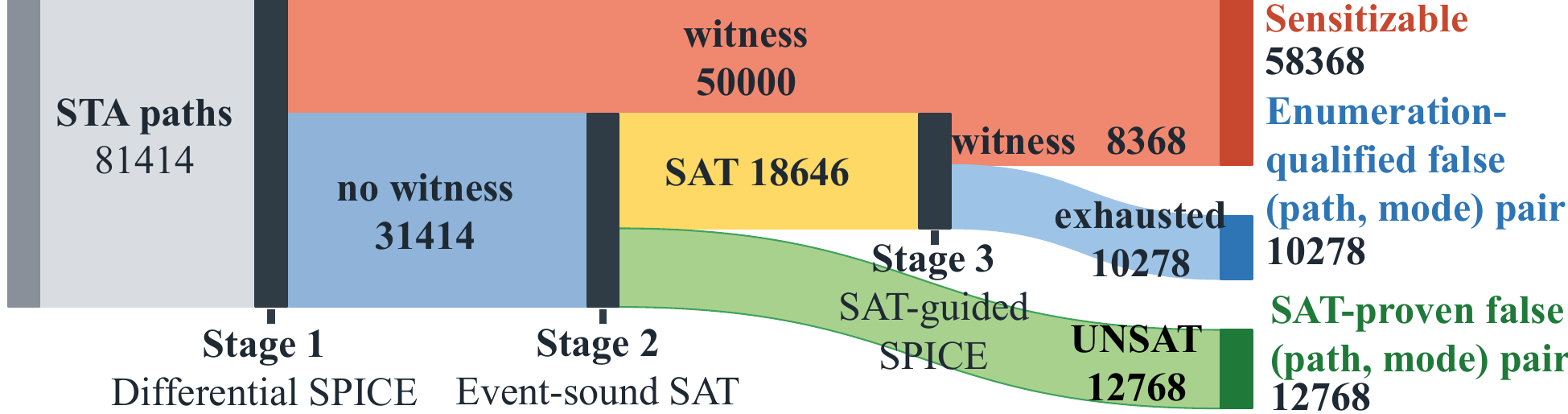}
    \vspace{-1em}
\caption{Progressive screening of STA-reported pairs on d05.}
    \label{fig:sankey}
    \vspace{-0.7em}
\end{figure}

Table~\ref{tab:false_path_results} summarizes the false-path identification results, where NanoTime extracts all timing paths. For both EvoPath and ESP, the classifications are identical across the three PVT corners on all designs, so we report the results per (path, mode) pair, with runtime averaged across corners. Fig.~\ref{fig:sankey} shows progressive screening of STA-reported pairs on d05, where only unresolved pairs proceed until all are classified.
The last two columns of Table~\ref{tab:false_path_results} report the pairs on which EvoPath and ESP disagree in false-path classification: those EvoPath reports false but ESP deems sensitizable, and those ESP reports false but EvoPath finds sensitizable. Two structural classes account for all disagreements.

The first class spans some designs with ESP comparison and involves paths
from sense enable to internal sense-amplifier nodes, contributing all
ESP-false / EvoPath-sensitizable disagreements in d01--d05. The root cause is a limitation inherent to switch-level abstraction when modeling analog-node behavior.
In SPICE, the internal sense node moves only slightly with the discharging
bitline before sense enable and transitions during regeneration afterward,
preserving the required event order. 
In switch-level model, the conducting transistor transfers the bitline logic
value to the sense node before sense enable, violating the event order and
causing a false classification.
Replaying the stimuli extracted from ESP in transistor-level SPICE
reproduces the sensitizable behavior, confirming that the disagreement
stems from the abstraction.

The second class, observed in d05, d06 and d08, arises in the column multiplexer. When the column address changes, timing skew between the outgoing and incoming column-select signals briefly enables two pass transistors simultaneously, causing charge sharing between a discharged and a precharged bitline. In SPICE, the resulting perturbation on the precharged bitline remains well below the characterized sensitization threshold, and EvoPath classifies these paths as false. The switch-level model, however, resolves the shared node to logic 0, interpreting the perturbation as a complete switching event. This artifact causes ESP to report 3840, 2066, and 2048 (path, mode) pairs in d05, d06, and d08, respectively, as sensitizable. All involve idle or read-port modes in which the affected bitline is not accessed. In d06 and d08, the same charge-sharing artifact resolves the selected read bitline to logic 0, masking its true state at the sense-amplifier input. ESP thus further classifies 378 and 512 clock-to-output read path pairs through the column multiplexer, respectively, as false, whereas EvoPath's SPICE witnesses show the read output switching.

The divergence direction is safety-critical. Classifying a sensitizable path as false removes a timing constraint and may mask a true violation, whereas reporting an unsensitizable path as sensitizable wastes engineering effort. In d01--d04, all ESP disagreements fall in the dangerous direction. The same artifact appears in other designs and produces divergences in both directions for some multi-port SRAMs. 
Every EvoPath result is established by a SPICE witness, an UNSAT proof, or
exhaustive SAT-guided enumeration.
These results expose a limitation of switch-level abstraction in representing analog-node behavior on transistor-level paths spanning digital and analog circuitry, making EvoPath a complementary approach for false-path identification and verification in this regime.

\subsubsection{Runtime Comparison}
EvoPath's runtime is dominated by differential SPICE simulation, with multi-port designs requiring more simulations due to larger mode-transition spaces. ESP is generally faster, using switch-level analysis and hierarchical compression. For d05 and d06, compression fails to preserve multi-port array behavior and is disabled, increasing runtime. Since ESP is a mature commercial tool and EvoPath a research prototype, runtimes are not directly comparable.

\vspace{-0.4em}
\subsection{Progressive Agentic Memory Evaluation}
\vspace{-0.1em}
\subsubsection{Memory Growth}
\begin{figure}[t]
    \centering
    \includegraphics[width=0.77\linewidth]{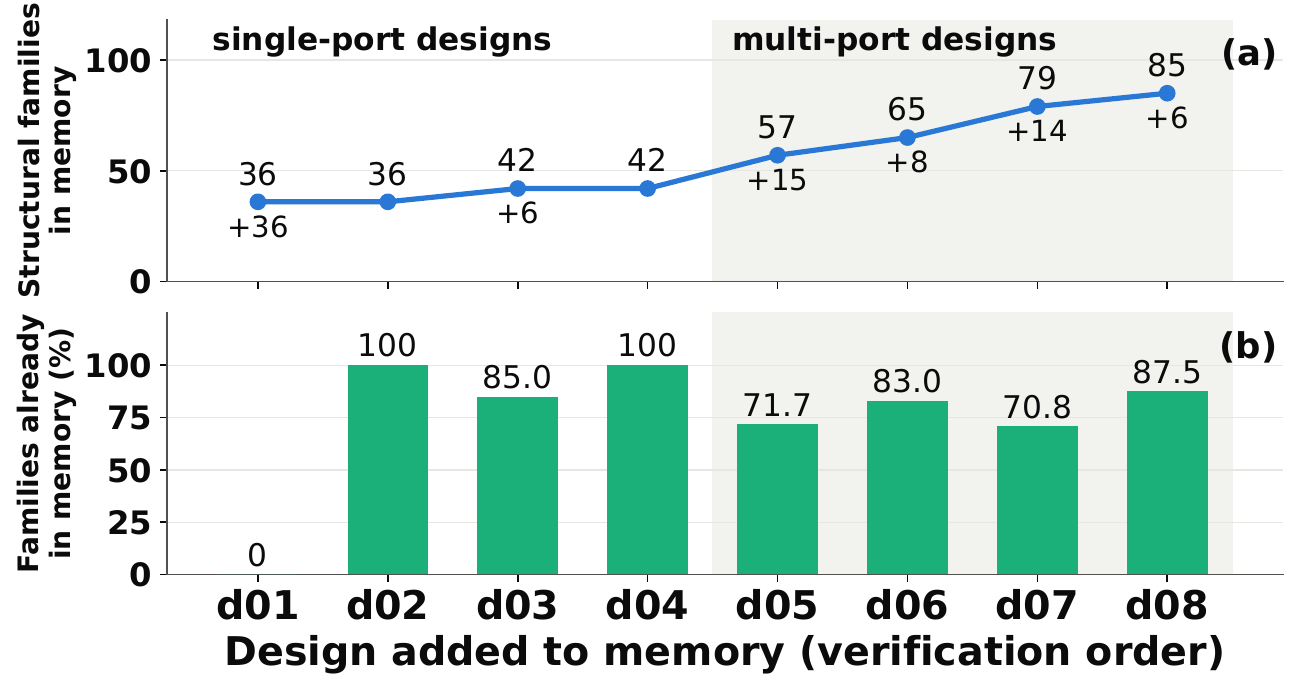}
    \vspace{-0.8em}
\caption{Memory growth across eight designs. (a) Cumulative and newly added structural families. (b) Fraction of each design's families already present in memory.}
    \label{fig:memory_growth}
    \vspace{-0.7em}
\end{figure}

% Fig.~\ref{fig:memory_growth} shows agentic memory growth across designs.
Fig.~\ref{fig:memory_growth}(a) reports the cumulative number of structural
families and new families from each design, while
Fig.~\ref{fig:memory_growth}(b) shows the fraction of each design's families
already in memory. Memory growth is driven by structural diversity
rather than design size. After the first single-port SRAM, varying capacity,
array organization, and column-mux ratio add few or no new families, as these
changes increase paths within existing families rather than creating new ones.
New port architectures introduce more unseen families, but subsequent designs
of the same architecture achieve high reuse. Memory therefore tends to saturate
within an architecture rather than growing with the number of designs.

% \subsubsection{Efficiency Gains from Memory Reuse}
\subsubsection{Memory Reuse}
We run EvoPath on d09–d12 with the family memory accumulated from d01–d08 and compare against the same runs with memory disabled. Classifications are identical in both settings. Prioritizing workloads that produced witnesses in earlier designs reduces total runtime by 32\%, 22\%, 27\%, and 30\%, respectively.

% \begin{table}[t]
% \centering
% \caption{Interactive mode on d09--d12. Memory denotes the family memory accumulated from d01--d08.}
% \vspace{-0.5em}
% \label{tab:interactive}
% \scriptsize
% \setlength{\tabcolsep}{3pt}
% \renewcommand{\arraystretch}{0.5}
% \resizebox{1\columnwidth}{!}{%
% \begin{tabular}{|c|c|c|c|c|}
% \hline
% \multirow{2}{*}{Design} & \multicolumn{2}{c|}{Hypothesis accuracy (\%)} & \multicolumn{2}{c|}{Agent tokens (k)} \\
% \cline{2-5}
%  & w/o memory & w/ memory & w/o memory & w/ memory \\
% \hline
% d09 & [x] & [x] & [x] & [x] \\
% d10 & [x] & [x] & [x] & [x] \\
% d11 & [x] & [x] & [x] & [x] \\
% d12 & [x] & [x] & [x] & [x] \\
% \hline
% \end{tabular}%
% }
% \vspace{-1em}
% \end{table}
\subsubsection{Interactive Path Exploration}
We evaluate interactive mode on d09--d12 using 20 designer queries per design.
Each query specifies a functional source and SRAM mode and asks the agent to
explore for a candidate false path. The agent localizes a candidate
in the graph views, predicts whether it is false, and submits it for
verification. Hypothesis accuracy is the fraction of predictions matching
verified results. Each design is evaluated with and without the family memory
accumulated from d01--d08.
All candidate paths are successfully localized in the CCB graphs. Hypothesis
accuracy without memory is 85\%, 80\%, 75\%, and 75\% on d09--d12, and rises
to 95\%, 95\%, 95\%, and 85\% with memory, as matched families provide
verified experience. Relative to no memory, agent-token consumption
decreases by 22\%, 28\%, and 35\% on d09, d10, and d12, respectively, and
changes little on d11. Importantly, memory only guides exploration and
hypothesis generation. Every candidate is still resolved by the same
verification flow.

\section{Conclusion}
We have presented EvoPath, an agentic framework for false-path identification
in transistor-level SRAM timing sign-off. An agent with graph-based circuit
understanding and domain skills recovers SRAM intent,
while every timing exception is established by deterministic SAT or SPICE
evidence. Verified experience is distilled into family memory for reuse across
related SRAMs. On 12 SRAM designs, EvoPath agrees with switch-level symbolic analysis on digital paths and complements it where analog behavior lies beyond that abstraction, while memory reuse reduces effort on unseen designs and makes agent reasoning cheaper and more accurate.

\IEEEtriggeratref{23} 
\bibliographystyle{myIEEEtran}
\bibliography{reference_brief}
\end{document}